\documentclass[aps,nofootinbib,showpacs,preprintnumbers,amsmath,amssymb]{revtex4}
\usepackage{epsfig}
\usepackage{epsf}
\usepackage{amssymb}
\usepackage{amsmath}

\usepackage{natbib}

\begin{document}
\preprint{USM-TH-265}

\title{Modified Contour-Improved Perturbation Theory}


\author{Gorazd Cveti\v{c}$^a$}
  \email{gorazd.cvetic@usm.cl}
\author{Marcelo Loewe$^b$}
  \email{mloewe@fis.puc.cl}
\author{Cristian Martinez$^b$}
  \email{cxmartin@uc.cl}
\author{Cristi\'an Valenzuela$^b$}
  \email{cvalenzuela@fis.puc.cl}
\affiliation{
$^a$\,Depto.~de F\'{\i}sica and Centro Cient\'{\i}fico-Tecnol\'ogico de Valpara\'{\i}so, 
UTFSM, Valpara\'{\i}so, Chile\\
$^b$\,Facultad de F\'{\i}sica, Pontificia Universidad Cat\'olica de Chile, Casilla 306, 
Santiago 22, Chile}

\begin{abstract}

The semihadronic tau decay width allows a clean extraction of the strong coupling constant at low energies. 
We present a modification of the standard ``contour improved" method based on a derivative expansion of the Adler function. 
The new approach has some advantages compared to contour improved perturbation theory. 
The renormalization scale dependence is weaker by more than a factor two and
the last term of the expansion is reduced by about 10\%,
while the renormalization scheme dependence remains approximately equal.
The extracted QCD coupling  at the tau mass scale is by 2$\%$ lower than the ``contour improved" value.
We find $\alpha_s(M_Z^2)=0.1211\pm 0.0010$.

\end{abstract}

\maketitle

\section{Introduction}

Quantum Chromodynamics, QCD, is the theory of strong interactions,
being the quark and gluon fields its basic degrees of freedom.
It describes a rich variety of phenomena including confinement,
chiral symmetry breaking, binding of hadrons and asymptotic freedom.
For some observables a perturbative expansion in powers of QCD coupling $\alpha_s$ is possible,
if the relevant energy scale is bigger than the QCD scale $E_{\text{QCD}}\sim 1$ GeV.
The strong coupling is a fundamental parameter of the Standard Model and
its determination is relevant in itself and for the identification of new physics.
Extractions of the strong coupling constant from experiments
covering energy scales from $M_{\tau}= 1.78$ GeV to $\sim 200$ GeV
are consistent at a 1$\%$ level (at the $M_Z$ scale),
providing an impressive test of asymptotic freedom over an energy range of two orders of magnitude \cite{Bethke:2009jm}.

In this note we study the extraction of $\alpha_s$ from inclusive semihadronic tau decay or,
more precisely,
from the ratio of semihadronic to leptonic tau decay widths

\begin{equation}
 R_\tau =\frac{\Gamma(\tau\rightarrow\text{had}\: \nu_\tau\: (\gamma))}{\Gamma(\tau\rightarrow e^-\: \bar{\nu}_e\: \nu_\tau\: (\gamma))}.
\end{equation}

This observable offers a unique and clean way of testing perturbative QCD (pQCD) at low energies,
because of the relatively large mass of the heaviest lepton that allows us to use pQCD
and because of the inclusive character of the observable avoiding the complication of hadronization. 
The theoretical expression for $R_\tau$ is an integral in energy $\sqrt{s}$, from $\sqrt{s}= 0$ to $\sqrt{s}=M_\tau$,
of the discontinuity of the two-point $W$-channel correlation function, $\Pi(s)$, times a kernel.
Of course, due to the low energies involved and because $\Pi(s)$ is being evaluated at the physical cut,
perturbative QCD cannot directly be used.
However, thanks to the analytic properties of the exact $\Pi(s)$, 
$R_\tau$ can be expressed as a $s$-plane contour integral along the circle $|s|=M_\tau^2$, 
with the Adler function in the integrand \cite{Narison:1988ni,Braaten:1988ea,Braaten:1991qm,LeDiberder:1992fr}. 
This allows a perturbative evaluation of $\Pi(s)$ and  therefore of $R_\tau$,
because the scale involved $|s|=M_\tau^2$ corresponds to a small absolute value of $\alpha_s$ 
and the contributions from the physical cut are suppressed
 (see details below).

Experimentally we can separate final states with net strangeness from final states without net strangeness,
and within the latter, the vector (V) from the axialvector (A) channels:

\begin{equation}\label{rtau}
  R_\tau = R_{\tau}^{S}+ R_{\tau}^{V}+ R_{\tau}^{A}.
\end{equation} 
If we are interested in the extraction of $\alpha_s$ from $\tau$ decays it is convenient to use
the vector plus axial  part of (\ref{rtau}), $R_{\tau}^{V+A}= R_{\tau}^{V}+ R_{\tau}^{A}$,
instead of the V and A parts separately.
There are two main reasons for this.
Firstly, the experimental separation between V and A channels introduces an extra uncertainty not present in the sum, and secondly,
within the Operator Product Expansion (OPE) the experimentally extracted non-perturbative contribution, $\delta_{NP}$, and its associated uncertainty
are suppressed in V+A compared with the V or A channels taken separately,
due to a partial cancellation in the sum \cite{Davier:2008sk}.
The experimental value of $R_{\tau}^{V+A}$ is obtained from  $R_{\tau}$ and  $R_{\tau}^S$.
The completely inclusive hadronic quantity is obtained from the measured
leptonic branching ratios

\begin{equation}
R_\tau = \frac {1-{\cal B}_e-{\cal B}_\mu}{{\cal B}_e} = \frac {1}{{\cal B}_e}-1.9726 = 3.640 \pm 0.010,
\end{equation}
and the updated strange part is \cite{Davier:2008sk}

\begin{equation} \label{}
R_{\tau}^{S}  =  0.1615 \pm 0.0040.
\end{equation}
Using these two values we have

\begin{equation} \label{expvalues}
R_{\tau}^{V+A}  =  3.479 \pm 0.011.
\end{equation}
This is the main experimental input for the extraction of the strong coupling constant.
The extracted value of the $\overline {\rm MS}$ coupling constant
from $R_{\tau}^{V+A}$ in Contour-Improved perturbation theory (CIPT)
at N$^3$LO order is 

\begin{equation}
\alpha_s^\text{CI}(M_Z^2)=0.1217\pm 0.0017,
\end{equation}
after renormalization group (RG) evolution up to $Z_0$ scale.
We observe a tension when comparing with the world average obtained by Bethke \cite{Bethke:2009jm}

\begin{equation}
\alpha_s^\text{Bethke}(M_Z^2)=0.1184\pm 0.0007,
\end{equation}
dominated by the lattice QCD extraction from the HPQCD collaboration \cite{Davies:2008sw}

\begin{equation}
\alpha_s^\text{lattice}(M_Z^2)=0.1183\pm 0.0008.
\end{equation}
On the other hand,
when comparing with the value extracted from $Z_0$ decays, also at N$^3$LO order \cite{Baikov:2008jh},

\begin{equation}
\alpha_s^Z(M_Z^2)=0.1190\pm 0.0026,
\end{equation}
or with the value extracted from the $p_T$ dependence of the inclusive jet cross section
in $p\, \bar{p}$ collisions at $\sqrt{s}=1.96$ TeV,
the most precise determination from hadron-hadron colliders \cite{Abazov:2009nc},

\begin{equation}
\alpha_s^{p \bar{p}}(M_Z^2)=0.1161^{+0.0041}_{-0.0048}\; ,
\end{equation}
we observe agreement within the uncertainties
(with a higher alpha central value from tau decays).

After the publication \cite{Baikov:2008jh} of the $\alpha_s^4$ correction to $R_\tau$,
several groups have used this result for the extraction of $\alpha_s$
and condensates in the context of the OPE
\cite{Baikov:2008jh,Davier:2008sk,Beneke:2008ad,Maltman:2008nf,Narison:2009vy,Dominguez:2009yt}.
The values of $\alpha_s$ obtained from these groups are partly incompatible with each other \cite{Bethke:2009jm},
mainly due to the usage of either CIPT or Fixed Order Perturbation Theory (FOPT).
The groups also have differences in the way they treat non-perturbative contributions.

Alternatively, evaluation of low energy QCD observables can be
performed in the context of analytic QCD \cite{Shirkov:1997wi}
(anQCD; for reviews and further references see 
\cite{Shirkov:2006gv,Prosperi:2006hx,Cvetic:2008bn}).
Such approaches have a number of advantages.
For example, the running coupling $\alpha_s^{\rm an }(Q^2)$ is an
analytic function of $Q^2$ ($Q^2 \equiv - q^2$), 
without the Landau singularities
outside the negative semiaxis in the complex $Q^2$-plane.
Therefore, evaluated expressions for $\Pi(s=-Q^2)$
in anQCD lead to identical results for the integrals
eqns.~(\ref{rt1}) and (\ref{rt2}). This is not true in perturbative QCD,
due to Landau singularities of $\alpha_s(Q^2)$ on the positive $Q^2$-axis.
Some anQCD models \cite{Cvetic:2009wx,Cvetic:2006gc}  can reproduce the correct values of $R_{\tau}^{V+A}$
       by adjusting some low-energy free parameters, but then some other attractive
       features are lost.
However, other anQCD models, with otherwise attractive features,
have a tendency to give too low
values of $R_{\tau}^{V+A}$,  
cf.~Refs.~\cite{Milton:1997mi,Milton:2000fi,Geshkenbein:2001mn,Cvetic:2009wx}.

The main point of this work is to present an alternative evaluation of $R_\tau$ within pQCD.
We call this procedure {\it modified CIPT}.
Due to the relatively small energy involved, there are important effects coming from the manner we use the renormalization group,
as we know from the difference between CIPT and FOPT.
It turns out that in modified CIPT the extracted value of $\alpha_s(M_\tau^2)$ is 2$\%$ lower than in CIPT,
decreasing the difference between the value obtained from tau decays and the world average from \cite{Bethke:2009jm}.
Using as input the experimental values  $R_{\tau}^{V+A}$ (eq. (\ref{expvalues})) and $\delta_{NP}$ (see below),
we obtain in modified CIPT
 $\alpha_s^\text{mCI}(M_\tau^2)=0.341\pm 0.008$,
and in CIPT
 $\alpha_s^\text{CI}(M_\tau^2)=0.347\pm 0.015$.
Evolving the coupling up to the $Z_0$ scale we get, respectively,
 $\alpha_s^\text{mCI}(M_Z^2)=0.1211\pm 0.0010$  and
 $\alpha_s^\text{CI}(M_Z^2)=0.1217\pm 0.0017$.
The quoted uncertainties are total uncertainties.

The modification of CIPT is simple.
Instead of the usual power series expansion 
$a+ c_1\,  a^2  + c_2\,  a^3 + \ldots$ (where: $a\equiv\alpha_s/\pi$),
the Adler function is expressed by a nonpower series of the form
 $a+ \tilde{c}_1\,  \tilde{a}_2 + \tilde{c}_2 \, \tilde{a}_3 + \ldots$,
where $\tilde{a}_{n+1}$ are proportional to the $n$'th derivative of the coupling $a(Q^2)$
and $\tilde{c}_n$ are the new expansion coefficients.
Thus, the terms of the series are proportional to the coupling and its derivatives, 
i.e. proportional to the coupling, the $\beta$-function and derivatives of the $\beta$-function.
Therefore, it can be said that the  $\beta$-function plays in  modified CIPT a more central role than in CIPT.
This expansion in derivatives of $\alpha_s$ was introduced in \cite{Cvetic:2006gc}
in the context of skeleton-motivated expansion and analytic QCD.
It turns out that compared to CIPT the new expansion shows a lower RG dependence and hence 
a lower theoretical error within the method.

The total hadronic ratio in $e^+$ $e^-$ collisions,
 $R_{e^+ e^-}(s)$,
is another timelike observable which can be expressed in terms of the corresponding Adler function.
Using the new expansion for the Adler function
we present a new and simpler expression for the RG-improved $R_{e^+ e^-}(s)$
in terms of the new couplings $\tilde{\alpha}_n$.

In Section \ref{SectionGeneral} we review the standard evaluation of $R_\tau$.
Section \ref{SectionmCIPT} contains the main part of this note, here we present and study the new approach.
The uncertainty in the extraction of $\alpha_s$ from $R_{\tau}^{V+A}$ is discussed in Section \ref{SectionUncertainty},
while in Section \ref{Sectionee} a new expansion for the RG-improved $R_{e^+ e^-}(s)$ is presented.
Finally, the conclusions are given in Section \ref{SectionConclusions}.

\section{CIPT and FOPT}\label{SectionGeneral}

The semihadronic tau decay ratio can be expressed as \cite{Tsai:1971vv}

\begin{equation}\label{rt1}
 R_\tau= \int_0^{M_\tau^2} \frac{ds}{M_\tau^2}
\bigg(1-\frac{s}{M_\tau^2}\bigg)^2  \bigg(1+2 \frac{s}{M_\tau^2}\bigg)\frac{1}{\pi} \text{Im}\: \Pi(s),
\end{equation} 
where $\Pi(s)$ is the correlator of two $W$-channel currents
(for more detailed expressions see the reviews \cite{Davier:2005xq,Chetyrkin:1996ia})

\begin{equation}
\Pi(s) = |V_{ud}|^2 (\Pi_{ud}^V (s)+ \Pi_{ud}^A(s)) + |V_{us}|^2 (\Pi_{us}^V (s)+ \Pi_{us}^A(s)),
\end{equation} 
with $V_{ij}$ the elements of the Cabibbo-Kobayashi-Maskawa matrix.
Eq. (\ref{rt1}) cannot be evaluated directly using perturbative QCD due to the small energy involved in the integral.
However there is a way out.
By general arguments we know that
the exact function $\Pi(s)$ is analytic function in the whole $s$-complex plane,
excluding the physical region $s\geq 0$.
Therefore, using the Cauchy theorem we have

\begin{equation}\label{rt2}
 R_\tau=\frac{-1}{2\pi i} \oint\limits_{|s|=M_\tau^2}    \frac{d s}{M_\tau^2}
\bigg(1-\frac{s}{M_\tau^2}\bigg)^2  \bigg(1+2 \frac{s}{M_\tau^2}\bigg) \Pi(s),
\end{equation} 
and integrating by parts

\begin{equation}\label{rt3}
 R_\tau=\frac{1}{2\pi i} \oint\limits_{|x|=1}    \frac{dx}{x}
(1-x)^3  (1+x) \, \frac{1}{2}  D(-x M_\tau^2),
\end{equation} 
where the Adler function $D(Q^2)$, defined by
\begin{equation}\label{adler}
D(Q^2)= - Q^2\frac{d\Pi(-Q^2)}{dQ^2},
\end{equation} 
is an observable (the renormalization scheme dependent constant of $\Pi(s)$ is eliminated).
Note that in eqns. (\ref{rt2}) and (\ref{rt3}) the quantities $\Pi(s)$ and $D(Q^2)$ are
evaluated at an absolute value of energy square equal to $M_\tau^2$.
Therefore the absolute value of the complex running coupling constant is small enough
in order to perform a perturbative treatment.
The theoretical expression for the vector plus axial part of $R_\tau$ can be written
in the Operator Product Expansion framework as

\begin{equation}\label{rtcontrib}
 R_\tau^{V+A}= 3 |V_{ud}|^2 S_{ew} 
(1 +  \delta_0 + \delta_{ew}^{\prime} + \delta_2 + \delta_{NP}),
\end{equation} 
where the perturbative QCD correction, the central object of this article, is given by $\delta_0$.
$S_{ew}=1.0198\pm 0.0006$ \cite{Marciano:1988vm} and 
$\delta_{ew}^{\prime}=0.001\pm 0.001$ \cite{Braaten:1990ef} are electroweak corrections,
$\delta_2=(-4.3\pm2.0)\times10^{-4}$ are light quark masses effects,
$\delta_{NP}=(-5.9\pm 1.4)\times 10^{-3}$ \cite{Davier:2008sk} are non-perturbative contributions,\footnote{This value
is obtained in CIPT. We assume that $\delta_{NP}$ does not vary in modified CIPT.}
and $V_{ud} =0.97418\pm0.00027$ \cite{Amsler:2008zzb}.
From these values and eq. (\ref{expvalues}) we obtain the experimental value 

\begin{equation}\label{d0exp}
\delta_0=0.204 \pm 0.004,
\end{equation} 
the number from which we extract $\alpha_s$ using only pQCD.

The perturbative QCD contribution can be expressed as
\begin{equation}\label{delta0}
 \delta_0=\frac{1}{2\pi i} \oint\limits_{|x|=1}    \frac{dx}{x}
(1-x)^3  (1+x) \,  \hat{D}(-x M_\tau^2),
\end{equation} 
where the reduced (canonically normalized) Adler function $ \hat{D}(Q^2)$, defined from eqns. (\ref{rt3}),  (\ref{rtcontrib}), and  (\ref{delta0}), is the massless QCD perturbative (leading twist) contribution to

\begin{equation}\label{} \nonumber
\frac{D(Q^2)}{3 |V_{ud}|^2 S_{ew}} - 1\longrightarrow \hat{D}(Q^2).
\end{equation} 
Perturbatively and without considering quark masses the vector and axialvector contributions are equal.
The vector (or axial) Adler function is known at fourth order in QCD

\begin{equation}\label{dsinRG}
\hat{D}(Q^2) = \sum_{n=1}^4 \, a^n(\mu^2)
\sum_{m=0}^{n-1} c_{n,m} \log^m(Q^2/\mu^2),
\end{equation} 
where $a(Q^2)\equiv \alpha(Q^2)/\pi$, $\mu$ is the renormalization scale,
and $c_{n,m}$ the expansion coefficients.
Only the coefficients $c_{n,0}$ are independent.
By using the renormalization-group (RG) equation,
the coefficients $c_{n,m}$ with $m\geq 1$ can be obtained as linear combinations of the $c_{n',0}$ with $n'<n$,
with coefficients depending on the perturbative $\beta$-function.
The Adler function is itself an observable and its RG improved expression is

\begin{equation}\label{dRG}
\hat{D}^\text{RG}(Q^2) = \sum_{n=1}^4 \, c_{n,0}\, a^n(Q^2).
\end{equation}
The $\beta$-function coefficients are normalized as

\begin{equation}\label{betaFunct}
\frac{\partial a}{\partial\log \mu^2} = \beta(a)=
-(\beta_0\, a^2+\beta_1\, a^3+\beta_2\, a^4+\beta_3\, a^5).
\end{equation}
In Contour Improved Perturbation Theory (CIPT) 
the expression for $\delta_0$, eq. (\ref{delta0}),
is evaluated using expression (\ref{dRG}) for the Adler function,
where the behavior of the coupling $a(Q^2)$ is exactly dictated by eq. (\ref{betaFunct}).
Thus, the Adler function is RG improved along the contour (\ref{delta0}),
which is the right thing to do considering that $d(Q^2)$,
and therefore the integrand of eq. (\ref{delta0}), is an observable.
An alternative procedure is Fixed Order Perturbation Theory (FOPT)
where eq. (\ref{dsinRG}) is used in (\ref{delta0}),
choosing a unique renormalization scale $\mu^2=M_\tau^2$ along the contour.
The central values of the coupling extracted using these two methods deviate significantly from each other

\begin{eqnarray}
\alpha_s^\text{CI}(M_\tau^2) &=& 0.347, \\
\alpha_s^\text{FO}(M_\tau^2) &=& 0.326.
\end{eqnarray}
Another argument in favor of CIPT is the fact that this procedure
is much more stable under renormalization scale variations than FOPT \cite{Davier:2008sk}.\footnote{
For discussions and comparisons between both approaches see \cite{Beneke:2008ad,Davier:2008sk,Caprini:2009vf,DescotesGenon:2010cr}.}
We will not consider FOPT in the following.
From the discrepancy between the two methods we learn that in the extraction of $\alpha_s$ from $R_\tau$
there is an important effect coming from the manner the RG is used.
This motivates the next section.

Before presenting the derivative expansion,
let us make a further comment.
It is well known that near the Minkowskian semiaxis the perturbative Adler function (or $\Pi(s)$) does a bad job at 
reproducing the exact function due to the resonance structure. 
Therefore the factor $(1-x)^3$ in eq. (\ref{delta0})  is crucial for a clean evaluation of $R_\tau$,
because it suppresses the contribution from this problematic region.\footnote{Duality violations are studied in \cite{Cata:2008ru,GonzalezAlonso:2010rn}.}

\section{Modified CIPT}\label{SectionmCIPT}

\begin{figure}[t]
\centering
\begin{tabular}{cc}
\epsfig{file=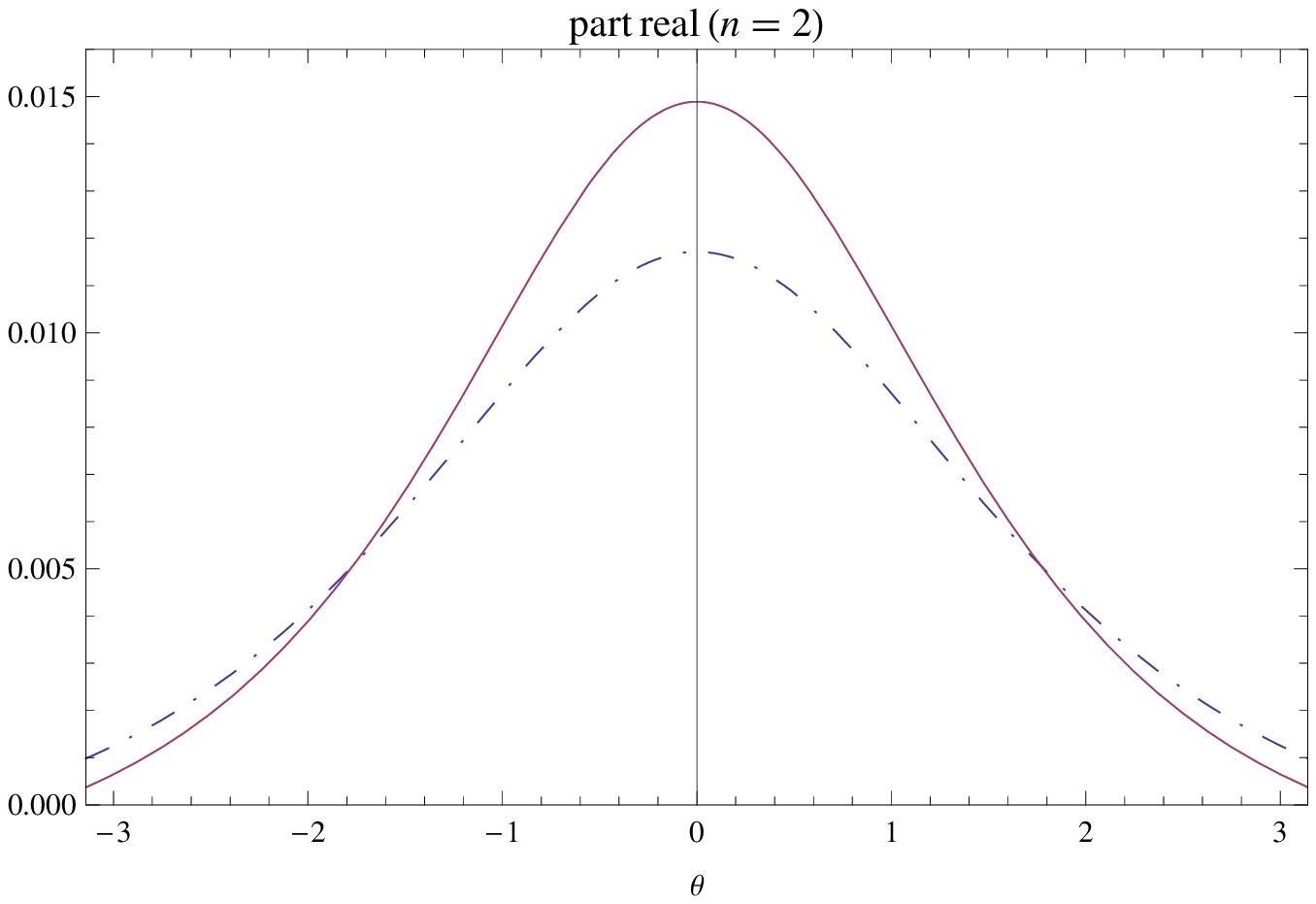,width=0.45\linewidth} &
\epsfig{file=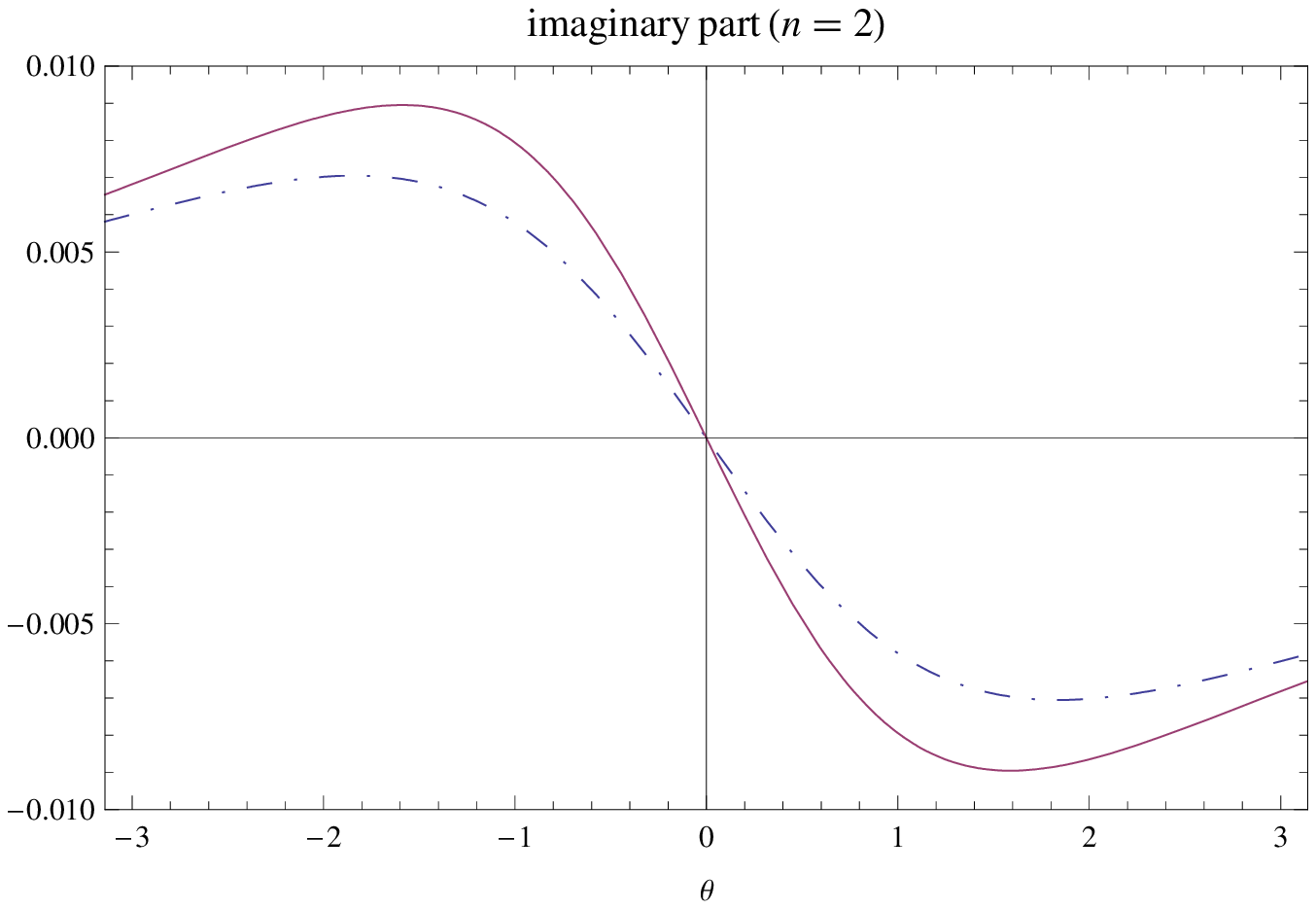,width=0.45\linewidth}
\end{tabular}
\vspace{-1cm}
\caption{\footnotesize Left: Real part of $\tilde{a}_2(M_\tau^2 e^{i \theta})$ (solid line)
as a function of $\theta$ compared to $a^2(M_\tau^2 e^{i \theta})$ (dashed line).
In both cases we take $a(M_\tau^2)= 0.340/\pi$.
Right: The corresponding imaginary part.}
\label{FigN2}
\end{figure}

\begin{figure}[t]
\centering
\begin{tabular}{cc}
\epsfig{file=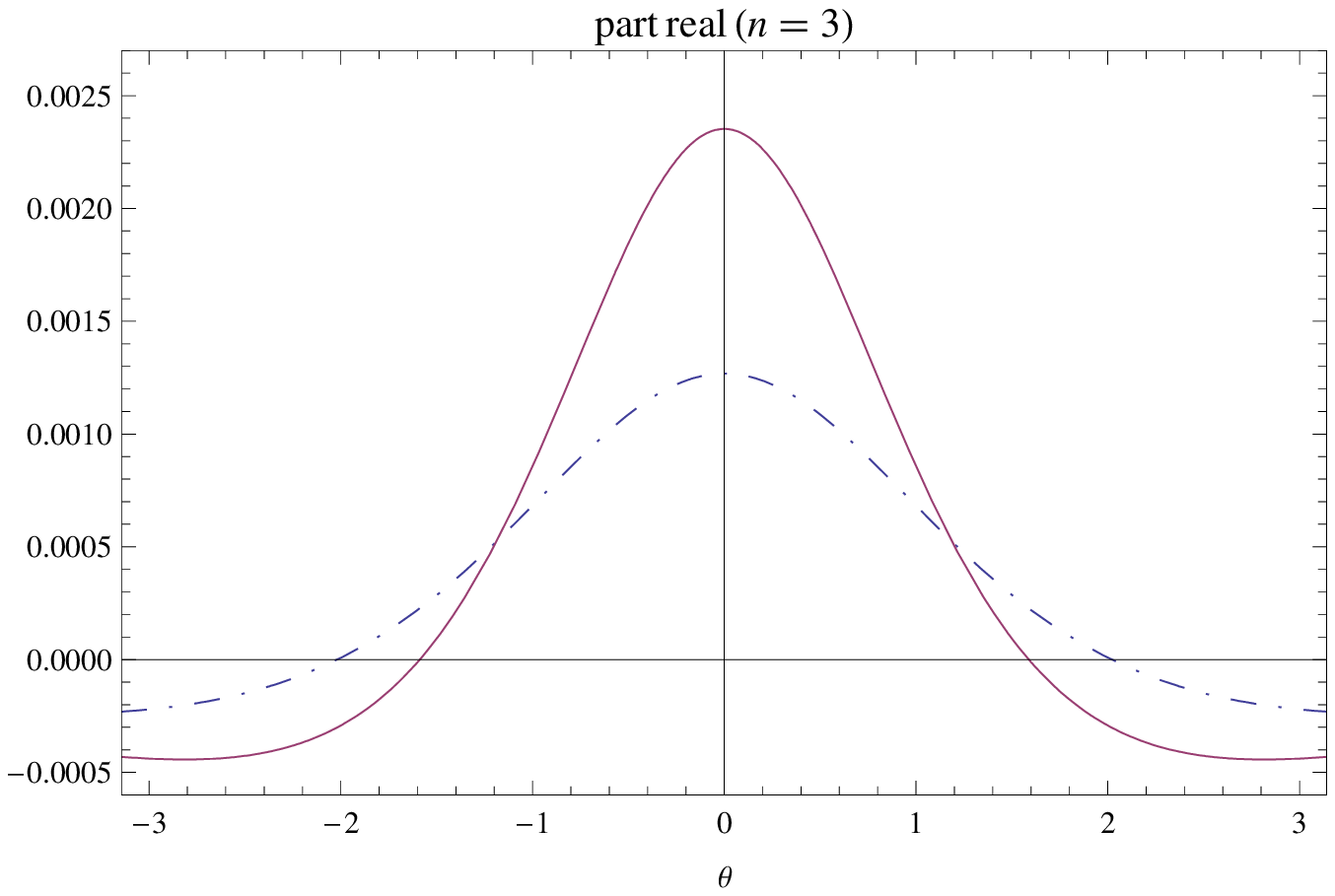,width=0.45\linewidth} &
\epsfig{file=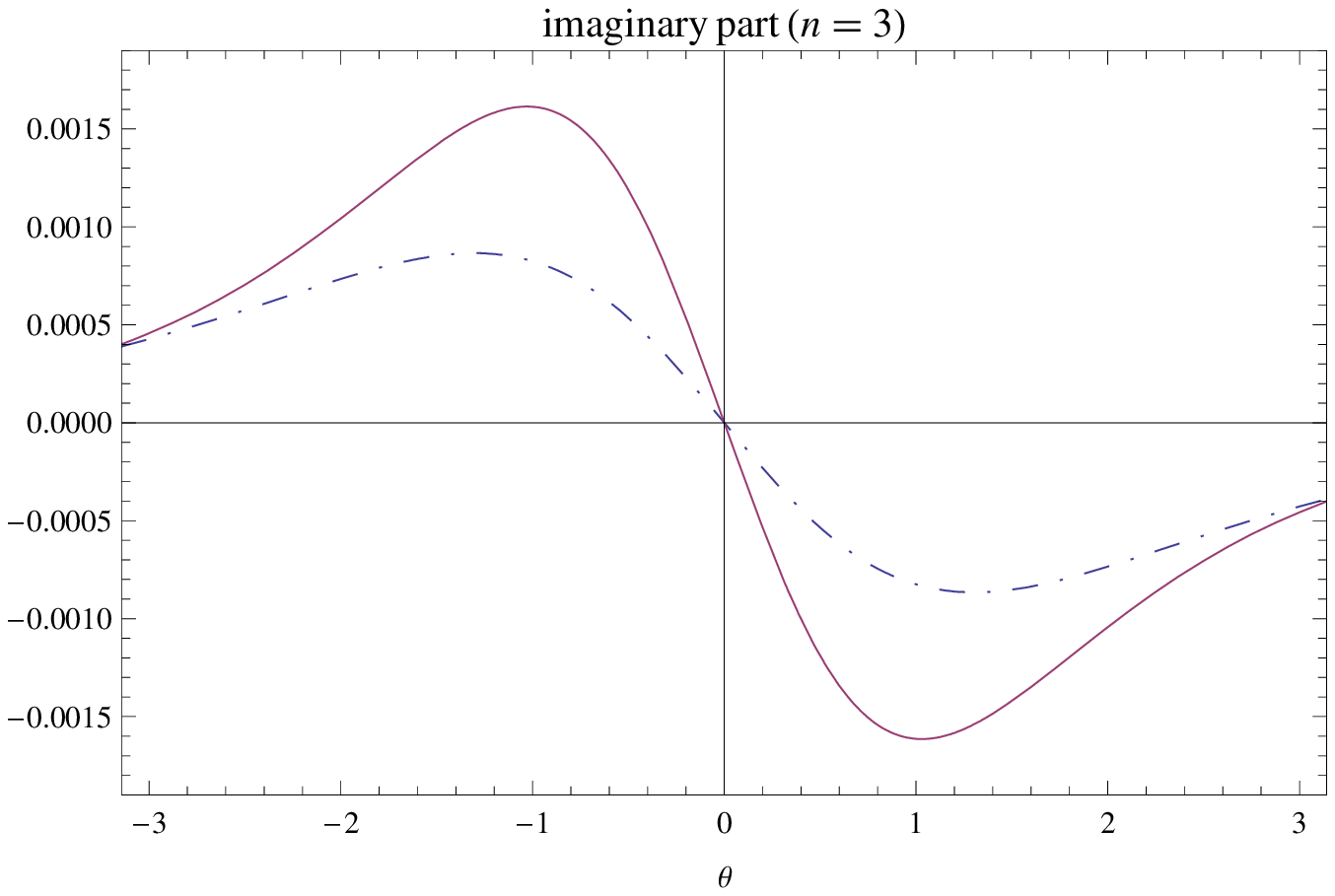,width=0.45\linewidth}
\end{tabular}
\vspace{-1cm}
\caption{\footnotesize Left: Real part of $\tilde{a}_3(M_\tau^2 e^{i \theta})$ (solid line)
as a function of $\theta$ compared to $a^3(M_\tau^2 e^{i \theta})$ (dashed line).
In both cases we take $a(M_\tau^2)= 0.340/\pi$.
Right: The corresponding imaginary part.}
\label{FigN3}
\end{figure}

\begin{figure}[t]
\centering
\begin{tabular}{cc}
\epsfig{file=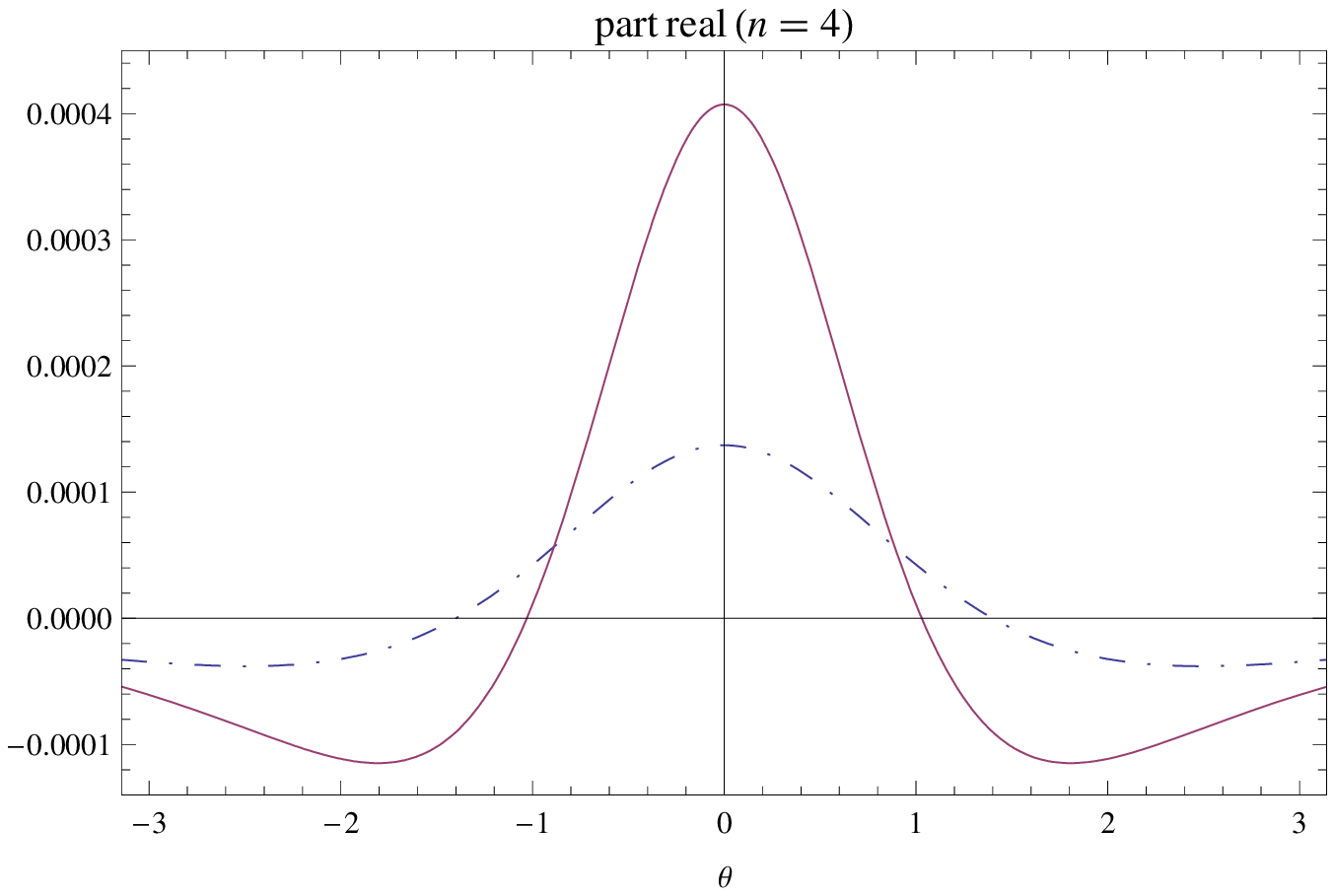,width=0.45\linewidth} &
\epsfig{file=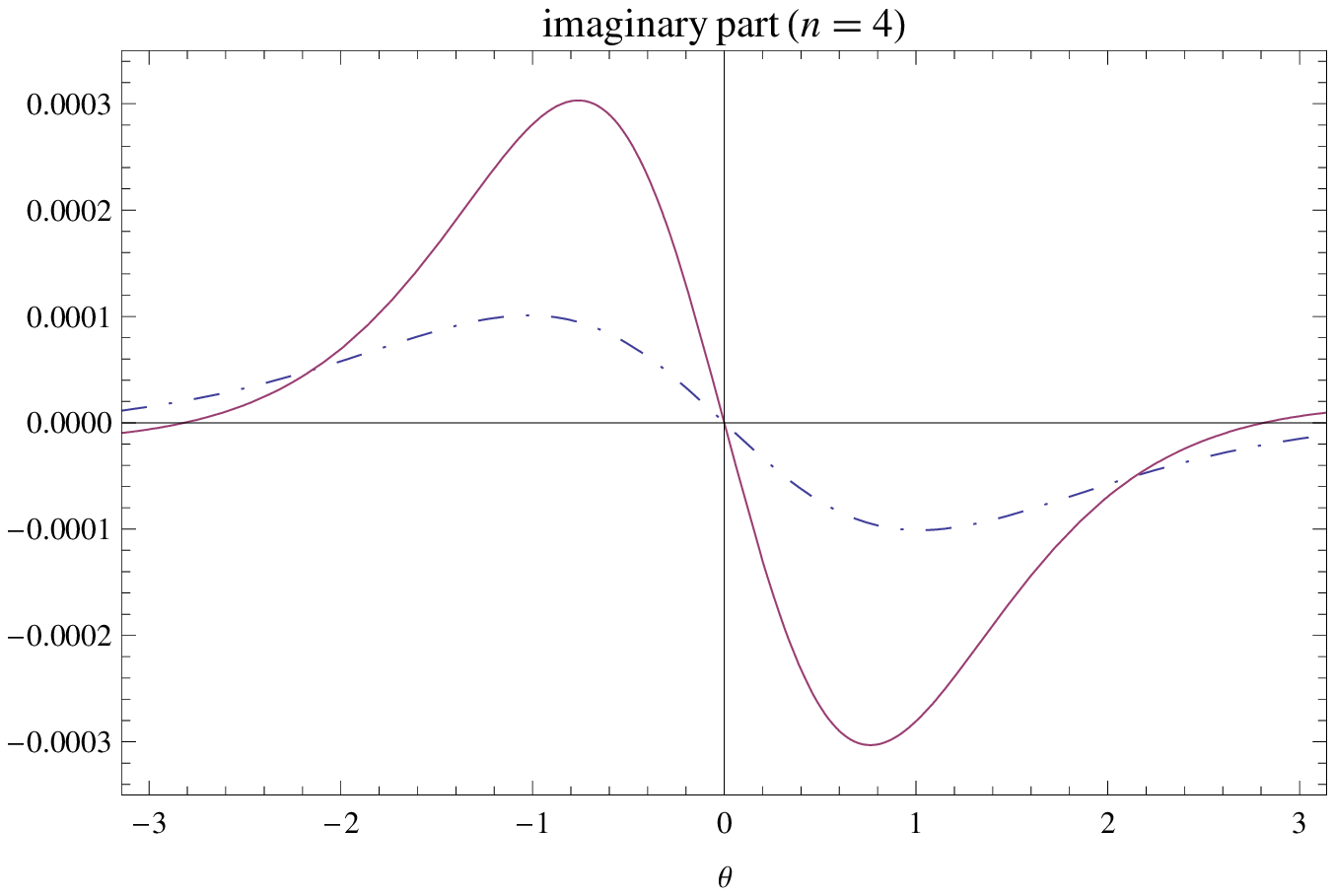,width=0.45\linewidth}
\end{tabular}
\vspace{-1cm}
\caption{\footnotesize Left: Real part of $\tilde{a}_4(M_\tau^2 e^{i \theta})$ (solid line)
as a function of $\theta$ compared to $a^4(M_\tau^2 e^{i \theta})$ (dashed line).
In both cases we take $a(M_\tau^2)= 0.340/\pi$.
Right: The corresponding imaginary part.}
\label{FigN4}
\end{figure}

In this, the central part of the article,
a new method for the evaluation of $R_\tau$ is presented.
We call the new approach modified CIPT.
As in CIPT, the semihadronic tau decay ratio is evaluated in eq. (\ref{delta0})
using for the Adler function $\hat{D}$ a RG-improved expression.
The proposed modification consists in using
instead of the standard series in powers of $a(Q^2)$ given in eq. (\ref{dRG})
a nonpower expansion for the Adler function in terms of the new {\it couplings} $\tilde{a}_n(Q^2)$.
Truncated at the last known term the new expansion is given by

\begin{equation}\label{dRGtilde}
\tilde{D}(Q^2) = \sum_{n=1}^4 \, \tilde{c}_n\, \tilde{a}_n(Q^2),
\end{equation}
with the tilde couplings defined as

\begin{equation}
  \tilde{a}_{m+1}      \equiv  \frac{(-1)^m}{\beta_0^m \, m!} \frac{d^m a}{d(\log Q^2)^m}.  \label{tildeDef}
\end{equation}
The derivatives are evaluated perturbatively using eq. (\ref{betaFunct}).
The new coefficients $\tilde{c}_n$ are obtained from the coefficients $c_{m,0}$ with $m\le n$.
The tilde couplings are normalized such that $\tilde{a}_n = a^n + \mathcal{O}(a^{n+1})$.
Note that in the new expansion $\tilde{a}_1 = a$,
$\tilde{a}_{2}=-\beta(a)/\beta_0$, 
$\tilde{a}_{3}=\beta(a)\beta'(a)/(2\beta_0^2)$, etc.
The beta function and its derivatives play in (\ref{dRGtilde}) a more central role than in eq. (\ref{dRG}).
The real and imaginary parts of the first three new couplings, $\tilde{a}_2$, $\tilde{a}_3$ and $\tilde{a}_4$, are plotted
along the integration circle of eq. (\ref{delta0}) 
in Figs.~\ref{FigN2}, \ref{FigN3} and \ref{FigN4}, respectively,
together with the corresponding powers $a^n$.
Note that in all three cases $\tilde{a}_n(M_\tau^2) > a^n(M_\tau^2)$.
In general, the ratio $\tilde{a}_n/a^n$ growths with $n$:

\begin{equation}
\frac{\tilde{a}_n}{a^n} =1+r_n+ \mathcal{O}(a^2),
\end{equation}
where
\begin{equation}
r_{n+1}=r_n (n+1)/n+\beta_1/\beta_0.
\end{equation}
Only for low values of $a$ and $n$ we have $\tilde{a}_n/a^n$ near 1.
The analyticity properties of the tilde and standard couplings are similar.
If we restrict ourselves to real $Q^2$, there are poles and cuts in the infrared region.

The series (\ref{dRG}) and (\ref{dRGtilde}) are formally equal if 
infinite number of terms in both expressions is considered.
However, it is believed that these series are asymptotic.
If we had the complete perturbative series we would truncate them at their respective (in absolute value) smallest terms
in order to give a meaning to the sums. 
In each case, the difference between the sum and the exact value of the original quantity
is expected to be smaller than the last considered term.
As a consequence, 
the rearrangement or reshuffling of eq. (\ref{dRG}) in eq. (\ref{dRGtilde}) is not immaterial:
it leads in general to different values of the 
truncated series for the Adler function.
Furthermore, with the presently known coefficients the asymptotic behavior of the Adler function is possibly not reached and
we are forced to truncate the series before, 
obtaining also different values in CIPT and modified CIPT.

Next we evaluate $R_\tau$ and compare CIPT and modified CIPT.
The ${\overline {\rm MS}}$ three-flavor Adler function coefficients $c_{n,0}$ and  $\tilde{c}_n$ are

\begin{eqnarray}
c_{1,0}&=&1,\; c_{2,0}= 1.6398,\; c_{3,0}=6.3710,\; c_{4,0}= 49.076,  \label{coefficients} \\
\tilde{c}_1&=&1,\quad \tilde{c}_2 =1.6398,\quad \tilde{c}_3 =3.4558, \quad\tilde{c}_4 =26.385.
\end{eqnarray} 
The first two coefficients are equal by construction
and the third and fourth tilde coefficients are about half the standard ones.
The value of the first unknown coefficient has been estimated \cite{Baikov:2008jh} 
using the method of ``fastest apparent convergence'' to be $c_{5,0}= 275$,
corresponding to $\tilde{c}_5=-25.4$.\footnote{If the unknown coefficient is estimated as
$c_{5,0}=c_{4,0}(c_{4,0}/c_{3,0})=378$, we obtain $\tilde{c}_5=+77.6$}

\begin{table}[h]  
\begin{ruledtabular}
\begin{tabular}{c|ccccc|cc|c}
$\hat{D},\tilde{D}$         &  1      &  2       &  3      &    4      &   5       &   $\sum_{n=1}^4$  & $\sum_{n=1}^5$ &   $a$    \\ \hline 
 $c_{n,0}\, a^n$            & 0.1132  & 0.0210   & 0.0092  &   0.0081  & (0.0051)  &  0.1515         & (0.1566)     &  0.3556/$\pi$ \\
$\tilde{c}_n\, \tilde{a}_n$ & 0.1082  & 0.0244   & 0.0081  &   0.0107  & (-0.0019) &  0.1515         & (0.1496)     &  0.3400/$\pi$ \\ 
\end{tabular}
\caption{Contributions to the Adler function using the standard power expansion (first row) and the tilde expansion (second row). 
In both cases the coupling $a$ is chosen in order to obtain $D(M_\tau^2)=0.1515$ when summing up to  $n=4$. 
Numbers in brackets consider the estimation for the first unknown coefficient, $c_{5,0}=275$.
\label{table_Adler}}
\end{ruledtabular} 
\end{table}

As an intermediate step, we first study the Adler function at the scale $M_\tau$ using 
the power series eq. (\ref{dRG}) and the tilde expansion eq. (\ref{dRGtilde}).
The results are shown in Table ~\ref{table_Adler}.
We demand in both cases $D(M_\tau^2)=0.1515$ when summing up to  $n=4$,
extracting a different value of strong coupling $a(M_\tau^2)$ in each case.
We observe in modified CIPT a value of the coupling $4-5\%$ smaller than in CIPT,
this difference is much smaller than the experimental error for the Adler function \cite{Eidelman:1998vc}.
The small ratio $\tilde{c}_n/c_{n,0}$ for $n=3$ and 4 is compensated
at this scale by a high ratio $\tilde{a}_n/a^n$.
If we take the estimate  $c_{5,0}= 275$  (see the numbers in brackets)
then this term is suppressed in the tilde expansion.

Note also that the $n=4$ contribution is close to the $n=3$ one.
This fact could be accidental or can be interpreted as the term $n=3$ (or 4) being the smallest term
of the asymptotic series for the Adler function.
In the latter case, for the best estimation of the Adler function we must truncate the series at this term.
However, the tilde expansion for the Adler function is thought as an intermediate step in the evaluation of $R_\tau$
or related quantities.

\begin{table}[h]  
\begin{ruledtabular}
\begin{tabular}{c|ccccc|cc|c}
  $\delta_0$           &  1      &  2       &  3      &    4      &    5      & $\sum_{n=1}^4$ & $\sum_{n=1}^5$ &   $a$         \\ \hline 
$\text{CI}$           & 0.1513  & 0.0308   & 0.0128  &   0.0090  &  (0.0038) &  0.2038        & (0.2077)       &  0.347/$\pi$ \\
$\overline {\text CI}$ & 0.1484  & 0.0372   & 0.0104  &   0.0078  & (-0.0001) &  0.2039        & (0.2037)       &  0.341/$\pi$ 
\end{tabular}
\caption{Contributions to the semihadronic tau decay width using CIPT (first row) and modified CIPT (second row). 
In both cases the coupling $a(M_\tau)$ is chosen in order to obtain $\delta_0=0.204$ when summing up to  $n=4$. 
Numbers in brackets consider the estimation for the first unknown coefficient, $c_{5,0}=275$.
\label{table_delta0_MSbar}}
\end{ruledtabular}
\end{table}

We evaluate $\delta_0$ in $\overline {\rm MS}$ scheme using the expression (\ref{delta0}), eq. (\ref{dRG}) for CIPT (CI), 
and eq. (\ref{dRGtilde}) for modified CIPT ($\overline {\text CI}$).
The results are shown in Table ~\ref{table_delta0_MSbar}.
Again, the value of $a(M_\tau)$ is taken different in the two approaches
in order to obtain the same value of $R_\tau$,
when summing up to $n=4$, $\delta_0=0.204$.
In modified CIPT we get a value of the strong coupling at the $\tau$ mass scale by about $2\%$ lower than in CIPT.
In addition, we observe in modified CIPT compared to CIPT a smaller last term of the series ($n=4$ term).
We see no signal of having reached the smallest term 
of the asymptotic series for $R_\tau$.
Including the estimate $c_{5,0}=275$ the modified CIPT series has a surprisingly low last term.

\begin{table}[h]  
\begin{tabular}{p{2cm}|p{2cm}|p{2cm}|p{2cm}}
\hline \hline
\hspace{5mm} $\xi$ &  \hspace{3mm} $a(\xi M_\tau^2)$  &   \hspace{3mm} $\delta_0,\;{\text CI}$  &  \hspace{3mm}$\delta_0,\overline {\text CI}$ \\
\hline
\hspace{5mm} 0.7   &  \hspace{3mm}  0.3831/$\pi$      &  \hspace{3mm}    0.2009                 &  \hspace{3mm}     0.2020                        \\      
\hspace{5mm} 1     &  \hspace{3mm}  0.3400/$\pi$      &  \hspace{3mm}    0.1984                 & \hspace{3mm}      0.2031                        \\
\hspace{5mm} 2     &  \hspace{3mm}  0.2812/$\pi$      &   \hspace{3mm}   0.1907                 &  \hspace{3mm}     0.1991                        \\
\hline \hline
\end{tabular}
\caption{Renormalization scale dependence of $\delta_0$. 
The absolute value of the (squared) renormalization scale is chosen to be $\mu^2=\xi M_\tau^2$ with $\xi=0.7$, 1, and 2.
We take as reference $a(M_\tau^2)=0.34/\pi$.
\label{table_RenScale} }
\end{table}

We study the renormalization scale dependence of $\delta_0$ in $\overline {\rm MS}$ scheme,
comparing CIPT and modified CIPT.
The (squared) renormalization scale is chosen to be $\mu^2=\xi M_\tau^2$ with $\xi=0.7$, 1, and 2,
and we use as reference $a(M_\tau^2)=0.34/\pi$.
The results for $\delta_0$ are shown in Table ~\ref{table_RenScale}.
Taking as a measure of the scale dependence of $\delta_0$ its range of variation when $\xi$ varies between 0.7 and 2,
we obtain 0.0102 and 0.0040, for CIPT and modified CIPT, respectively.
If we extract $\alpha_s$ from the experimental value of $\delta_0$, these uncertainties translate to  
$\Delta \alpha(M_\tau^2)=$0.013 and 0.005 for CIPT and modified CIPT, respectively.
Thus, using this criterion, renormalization scale dependence is by more than factor two weaker in modified CIPT
than in the standard CIPT.

\begin{table}[h]  
\begin{ruledtabular}
\begin{tabular}{c|ccccc|cc|c|c}
$\delta_0$        &  1      &  2       &  3      &    4      &    5      &$\sum_{n=1}^4$ & $\sum_{n=1}^5$ &  $a({\rm tH})$  & $a\equiv a(\overline {\rm MS})$  \\ \hline 
$\quad {\text CI}$           & 0.1427  & 0.0284   & 0.0200  &   0.0129  & (0.0060)  &  0.2040     & (0.2100)     &  0.32908/$\pi$  &  0.3514/$\pi$\\
$\quad \overline {\text CI}$ & 0.1400  & 0.0326   & 0.0203  &   0.0112  & (0.0021)  & 0.2040      & (0.2061)     &  0.32354/$\pi$  &  0.3446/$\pi$
\end{tabular}
\caption{Renormalization scheme dependence of $\delta_0$. 
The quantity $\delta_0$ is evaluated in the 't Hooft scheme 
extracting $a({\rm tH})$ in CIPT and modified CIPT.
In the last column the value of $a({\rm tH})$ in converted to the $\overline {\rm MS}$ scheme.
\label{table_RenScheme} }
\end{ruledtabular}
\end{table}

Finally, we study the renormalization scheme dependence in the extraction of $\alpha_s$ from $\delta_0$, 
comparing $\overline {\rm MS}$ and 't Hooft (tH) schemes.
The latter scheme is defined by setting all but the first two coefficients of the beta function equal zero, 
i.e. $\beta_2=\beta_3=\ldots=0$. 
The result of the extraction of $a$ in the 't Hooft scheme is shown in Table ~\ref{table_RenScheme}.
Comparing these values with the values of (\ref{table_delta0_MSbar}),
we obtain $\Delta \alpha(M_\tau^2)=0.004$ in both cases.
Thus, renormalization scheme dependence is approximately equal in CIPT and modified CIPT
(it is slightly weaker in modified CIPT).

Modified CIPT is a new kind of perturbative expansion.
The approach is valid in itself and  possesses some attractive properties
as is a lower renormalization scheme dependence.
However, from the point of view of the standard power series of the Adler function,
mCIPT performs the sum (\ref{dRG}) up to $n=8$
considering nonzero coefficients $c_{n,0}$ for $n=5$ to $8$.
For example, 
when expanding the truncated expression (\ref{dRGtilde}) in powers of $a$,
we obtain $c_{5,0}=300.4$.
The natural question to ask is how good this value compares to the exact one.
To answer it we need to calculate the corresponding Feynman diagrams,
a task not expected to be done in the near future.
What we can do as a test of the method is to compare its prediction for 
the known coefficients $c_{3,0}$ and $c_{4,0}$.
From $c_{1,0}$ and $c_{2,0}$ the estimate for $c_{3,0}$ is 2.92,
and from $c_{1,0}$, $c_{2,0}$ and $c_{3,0}$ the estimate for $c_{4,0}$ is 22.7.
Comparing with the exact values, cf. eq. (\ref{coefficients}), 
we see that in both cases mCIPT 
includes a significant part of the next term of the power series.
Therefore, at least in these two cases,
mCIPT is an improvement also from this point of view.
Both estimates (for $c_{3,0}$ and $c_{4,0}$) are lower than the exact value by a factor 2.2 
(2.19 and 2.16 respectively).

\section{Uncertainty in the extraction of $\alpha_s$}\label{SectionUncertainty}

The experimental uncertainty in the extraction of $\alpha_s$,
$\Delta \alpha^{\text{exp}}=\pm 0.005$, comes from the
uncertainty in the value of the pseudo-observable quantity $\delta_0$ given in eq. (\ref{d0exp}).
By far the main contribution here is due to the experimental value of $R_{\tau}^{V+A}$, given in eq. (\ref{expvalues}).
The non-perturbative contribution $\delta_{NP}$ is considered here as an input for $\delta_0$
and also contributes to the experimental uncertainty of $\alpha_s$.
Its value and its associated uncertainty,
which are obtained from the moments of $R_{\tau}$,
are rather small   \cite{Davier:2008sk} and
the uncertainty in $\delta_{NP}$ has almost no relevance for the uncertainty of  $\delta_0$
(while the effect of $\delta_{NP}$ on the central value of $\delta_0$ is $~0.006$,
i.e. 1.5 times $\delta_0$'s experimental uncertainty).

The theoretical uncertainty, within modified CIPT, 
was obtained in the previous section varying the renormalization 
scale,  $\Delta \alpha^{\text{scl}}=0.005$, and 
scheme, $\Delta \alpha^{\text{sch}}=0.004$.
Adding them in quadrature we obtain $\Delta \alpha^{\text{theo}}=0.006$.
Then, the extracted value of the strong coupling constant at the $M_\tau$ scale in modified CIPT is

\begin{eqnarray}\label{alfatau}
\alpha_s^{\text{mCI}}(M_\tau^2) &=& 0.341 \pm 0.005^{\text{exp}} \pm 0.006^{\text{theo}}, \\
                                          &=& 0.341 \pm 0.008. \nonumber
\end{eqnarray}

For comparison we give the corresponding values in CIPT:

\begin{eqnarray}
\alpha_s^{\text{CI}}(M_\tau^2) &=& 0.347 \pm 0.005^{\text{exp}} \pm 0.014^{\text{theo}}, \\
                              &=& 0.347 \pm 0.015. \nonumber
\end{eqnarray}

Alternatively, we could estimate the theoretical uncertainty for $\alpha_s$ as coming uniquely from the way we use the RG:
taken the difference between the extracted central values using CIPT and modified CIPT.
Coincidentally, this would lead to the same theoretical uncertainty obtained above within modified CIPT.
We are not allowed to sum them. This would imply a double counting
because, by definition, the difference between CI and modified CI is given by higher order contributions.

Conventionally, we compare the values of the strong coupling extracted from different experiments
at a particular scale, the $M_Z$ scale. 
We perform the evolution at four loops, with three-loop matching conditions \cite{Chetyrkin:1997sg}
at the thresholds $\mu_{\rm thr} = 2 m_q$ ($q=c, b$).
The uncertainty $\pm 0.0005$ due to RGE evolution is as given in \cite{Davier:2008sk}. 
Evolving eq. (\ref{alfatau}) from $M_\tau$ to $M_Z$ we get 

\begin{eqnarray}\label{alfaZ}
\alpha_s^{\text{mCI}}(M_Z^2)    &=& 0.1211 \pm 0.0006^{\text{exp}} \pm 0.0007^{\text{theo}} \pm 0.0005^{\text{evol}}, \\
                   &=& 0.1211 \pm 0.0010, \nonumber
\end{eqnarray}
 in modified CIPT.

\section{Hadronic ratio in $e^+\; e^-$ collisions}\label{Sectionee}

Another Adler function related quantity is the ratio of the total hadronic to muonic cross sections, $R_{e^+ e^-}(s)$,
which is proportional to $ \text{Im}\: \Pi(s)$ (with $\Pi(s)$ in the neutral channel).
Analogous to the definition of $\hat{D}(Q^2)$ a reduced  $R_{e^+ e^-}(s)$, $\hat{R}(s)$, can be defined.
The function $\hat{R}(s)$ can be written as a function of the reduced Adler function as

\begin{equation}\label{RfromD}
\hat{R}(s)=\frac{1}{2\pi i}
\int\limits_{-s-i\varepsilon}^{-s+i\varepsilon}
\frac{dz}{z}\hat{D}(z).
\end{equation}
The non-RG-improved expressions for $\hat{R}(s)$ and $\hat{D}(Q^2)$ differ in the so-called $\pi^2$-terms, which are numerically important.
A RG-improved expression for $\hat{R}(s)$ cannot be obtained directly from the non-RG-improved $\hat{R}(s)$,
because the RG is valid in the Euclidean region and $s$ is timelike.
The right way to proceed is to apply the RG group to the Adler function, obtaining $\hat{D}^\text{RG}(Q^2)$,
and then get an improved version of $\hat{R}(s)$ from eq. (\ref{RfromD}) \cite{Radyushkin:1982kg}.

Using the tilde expansion
we obtain a simple expression for $\hat{R}(s)$,
because the integral (\ref{RfromD}) can be performed for all but the first term of the tilde series of the Adler function:

\begin{equation}\label{Rtilde}
\tilde{R}(s)=\frac{1}{2\pi i}
\int\limits_{-s-i\varepsilon}^{-s+i\varepsilon}
\frac{dz}{z}a(z)
+\sum_{n=2}^4
\frac{(-\tilde{c}_n)}{(n-1)\beta_0 \pi}
\text{Im}\;
\{\tilde{a}_{n-1}(-s-i\varepsilon)\}.
\end{equation}
Thus, we obtain a new expression for $R_{e^+ e^-}(s)$.
The first term of eq. (\ref{Rtilde}) is the Minkowskian coupling \cite{Radyushkin:1982kg},
and higher terms of the series are proportional to the imaginary part of the tilde couplings
evaluated at the time momentum $s$, times a suppressed coefficient.
It would be interesting to perform a phenomenological analysis of  $R_{e^+ e^-}(s)$
using this new expression.

\section{Conclusions}\label{SectionConclusions}

In this note we present a modification of Contour-Improved Perturbation Theory,
i.e., of the standard method for the evaluation of 
the semihadronic tau decay width, $R_\tau$,
within the context of pQCD and the OPE. 
Due to the low energy involved the truncated $\alpha_s^4$ result is sensible to higher order terms.
The way we use the renormalization group in calculating $R_\tau$ from the Adler function
leads to important uncertainties in the evaluation of $R_\tau$,
as is well known from the difference between CIPT and FOPT.
Both, in CIPT and in the proposed approach, the Adler function is evaluated using a varying renormalization scale
and not a fixed one $\mu=M_\tau$ as in FOPT.
The new ingredient in modified CIPT is in the series we use to evaluate the Adler function:
instead of the usual power series expansion 
$a+ c_1\,  a^2  + c_2\,  a^3 + \ldots$ (where $a\equiv\alpha_s/\pi$),
the Adler function is expressed by a nonpower series of the form
 $a+ \tilde{c}_1\,  \tilde{a}_2 + \tilde{c}_2 \, \tilde{a}_3 + \ldots$,
where the new tilde couplings $\tilde{a}_{n+1}$ are proportional to the $n$'th derivative of the coupling $a(Q^2)$
and $\tilde{c}_n$ are the new expansion coefficients.
It can be said that the  $\beta$-function plays in  modified CIPT a more central role than in CIPT.
This expansion in derivatives of $\alpha_s$ was introduced in \cite{Cvetic:2006gc}
in the context of skeleton-motivated expansion and analytic QCD.

Modified CIPT has some advantages compared to contour improved perturbation theory. 
The renormalization scale dependence is weaker by more than a factor two and
the last term of the expansion is reduced by about 10\%,
while the renormalization scheme dependence remains approximately equal.

The total hadronic ratio in $e^+$ $e^-$ collisions,
 $R_{e^+ e^-}(s)$,
is another timelike observable which can be expressed in terms of the corresponding Adler function.
Using the new expansion for the Adler function
we present a new and simpler expression for the RG-improved $R_{e^+ e^-}(s)$
in terms of the new couplings $\tilde{\alpha}_n$.

The extracted value of $\alpha_s$ from the vector plus axial non-strange $R_\tau$
is in modified CIPT 1.8\% (0.5\%) lower than in CIPT,
at the $\tau$ ($Z_0$) scale.
We obtain  
$\alpha_s^{\text{mCI}}(M_\tau^2)=0.341 \pm 0.008$ and 
$\alpha_s^{\text{mCI}}(M_Z^2)   =0.1211\pm 0.0010$.

\section*{Acknowledgments}

M.L. and C.V. would like to thank Ces\'areo Dom\'{\i}nguez, and C.V. would like to thank Bogdan Malaescu for interesting discussions.
This work has been supported by
Conicyt (Chile) Bicentenario Project PBCT PSD73 (C.V.),
FONDECYT Grant No. 1095217 (M.L., C.V., C.M.), 
FONDECYT Grant No. 1095196 (G.C.), and 
Proyecto Anillos ACT119 (G.C., M.L.)

\bibliography{modCIPT}

\end{document}